\documentclass{article}[12pt]
\usepackage[dvips]{graphicx}

\begin{document}

\title{A constrained stochastic state selection method applied to quantum spin systems}

\author{Tomo Munehisa and Yasuko Munehisa}



\maketitle

\begin{abstract}
We describe a further development of the stochastic state selection method, 
which is a kind of Monte Carlo method we have proposed in order to numerically study
large quantum spin systems. 
In the stochastic state selection method 
we make a sampling which is simultaneous for many states. 
This feature enables us to modify the method so that 
a number of given constraints are satisfied in each sampling.
In this paper we discuss this modified stochastic state selection method
that will be called the {\em constrained stochastic state selection method} 
in distinction from the previously proposed one
(the conventional stochastic state selection method) in this paper.
We argue that in virtue of the constrained sampling  
some quantities obtained in each sampling become more reliable, {\it i.e.}
their statistical fluctuations are less than those 
from the conventional stochastic state selection method. 

In numerical calculations of the spin-1/2 quantum Heisenberg antiferromagnet 
on a 36-site triangular lattice we explicitly show that data errors
in our estimation of the ground state energy are reduced.  
Then we successfully evaluate several low-lying energy eigenvalues of the 
model on a 48-site lattice. 
Our results support that this system can be described by the theory based on the 
spontaneous symmetry breaking in the semi-classical N\'{e}el ordered antiferromagnet.
 
\end{abstract}


\section{Introduction}
\label{sec1}
It is widely recognized that 
numerical methods based on the first principle are quite important
in the study of quantum spin systems. 
Actually the quantum Monte Carlo method has contributed toward enlarging our 
knowledge of non-frustrated quantum spin systems, especially of
the spin-1/2 quantum Heisenberg antiferromagnet on bipartite 
lattices\cite{book1,book2,rev}. 
Yet it is well known that this Monte Carlo method has a limited power
in calculations of two-dimensional frustrated spin systems.
In this situation many studies on numerical methods have been
proposed and developed, sometimes depending on approximations. 
One of these studies is the coupled-cluster method\cite{ccm1,ccm2}, 
which is a kind of variational method. 
Another approach is the stochastic reconfiguration method\cite{Sorella} 
whose origin is assigned to the fixed-node Monte Carlo method\cite{fixednode2,fixednode}.
One should also attend to various studies on the density matrix renormalization 
group (DMRG) method\cite{Hene,Mae,White3} which extend the original DMRG method on a 
chain\cite{White1, White2} to higher dimensional systems.
The path-integral renormalization group method\cite{IK} is interesting as well. 

Recently we have developed another Monte Carlo method, 
which we call the stochastic state selection (SSS) 
method\cite{MM1,MMss,MM2,MM3,MMtri,MMeq}.  This method has a good
property common to the ordinary Monte Carlo method that in principle
one does not need any approximations specific to the system we investigate.
The sampling algorithm is, however, quite different from the ordinary one
since the SSS method is based on 
not importance sampling but a new type of stochastic selection. 
In the SSS method we use an operator which generates sampled states  
from any given state. This operator includes 
a set of stochastic variables which are as many as the number of basis states 
of the vector space under consideration. 
The essential point of the selection is that 
many of these stochastic variables are valued to be zero while 
their statistical averages are all equal to one.
Therefore in this algorithm we select   
a relatively small number of basis states from a vast vector space 
in a mathematically justified manner so that
the statistical averaging processes give us the correct value of any inner product.
So far all these stochastic variables have been generated independently.
 
In this paper we propose the {\em constrained SSS method} --- a modified SSS method whose 
samples satisfy a number of given constraints.
In order to make it possible we introduce some dependencies among 
the stochastic variables stated above. 
From theoretical point of view all we need to restore the original state from sampled states is that the statistical average of each stochastic variable is equal to one.
Therefore we can define some variables in the set as functions of other variables,
instead of requesting they should be independently generated, if each of these functions assures that the statistical average of the dependent variable equals to one.
Keeping this in mind we develop the constrained SSS method, where values of special inner products to represent constraints are unchanged in each sampling.
We then argue that sampling errors in numerical studies can be reduced with suitable constraints and with suitable dependent variables.          

As a concrete example, we calculate energy eigenvalues around the ground state  
of the spin-1/2 quantum Heisenberg antiferromagnet on an $N_s$-site triangular lattice. 
The Hamiltonian of the system is 
\begin{equation}
\hat H = J_{\triangle} \sum_{(i,j)} \vec{S}_i  \cdot \vec{S}_j \,,
\label{hamil}
\end{equation}
where $\vec{S_i}$ denotes the spin-1/2 operator on the $i$-th site  
and the sum runs over all $N_b(=3N_s)$ bonds of the lattice. 
The coupling $J_\triangle$ is set to 1 throughout this paper.
In our calculations we employ the power method. 
The constrained SSS method is used to calculate the expectation values of powers of the operator 
$\hat Q $,
\begin{equation} 
\hat Q \equiv l \hat I - \hat H ,
\label{qdef}
\end{equation}
where $\hat I$ denotes the identity operator 
and $l$ is a positive number which depends on the lattice size. 
For each lattice size we choose one value of $l$ which ensures that the ground state eigenvalue of the system corresponds to the eigenvalue of $\hat Q$ whose absolute value is the largest.
Detailed explanations for it are given in Appendix A.
 
One reason why we study this system here is that, as is well-known,
it is a typical example of strongly frustrated systems in two dimensions.
Another reason is that there has been a long history of investigations into what 
state is realized on the triangular lattice.
Lots of studies on this
system\cite{rev,ccm1,ccm2,White3,Singh,Leung,Sindz,Bernu,Boninsegni,Capri,Weihong,
Trumper,Sorella2,Sorella1}
indicate the ground state with the three-sublattice order\cite{Singh}. 
One should, however, 
also note recent works\cite{Fendley,Oleg,Zheng} which suggest that this quantum system
has a richer phase structure than the one expected from the classical spin wave theory, 
as well as other studies\cite{ccm1,Sorella1} which show that this system is near 
the quantum critical point.

The plan of this paper is as follows. 
In the next section we describe the method. 
Subsection \ref{subsec21} is devoted to a brief review of the conventional SSS 
method\cite{MM2}.    
In subsections \ref{subsec22} and \ref{subsec23} 
we make a detailed account of the constrained SSS method. 
First we give a simple example to explain how we impose a constraint in a sampling
in subsection \ref{subsec22}.
Then extensions to more general cases are discussed in subsection \ref{subsec23}.
Sections \ref{secexact}, \ref{secappx} and \ref{sec48site} 
are for applications of the method 
to the spin 1/2 quantum Heisenberg antiferromagnet on triangular lattices. 
In section \ref{secexact} we study the model on a $16$-site lattice, 
where the exact eigenstate can be easily obtained.
Using this exact eigenstate we evaluate expectation values of the $m$-th power of the 
operator $\hat Q$ by means of the constrained SSS method.
We find that our results from one sampling coincide with the exact 
expectation values.
In section \ref{secappx} we investigate the model on the $36$-site lattice, for which 
a number of low-lying energy eigenvalues are known  
from the exact diagonalizations\cite{Bernu}. 
Starting with approximate states for the ground state of the system, 
which we obtain through procedures described in Appendix B,
we argue the accuracy of the constrained SSS method by comparing our results 
with the ground state energy reported in ref.\cite{Bernu}.
The resultant expectation values show that the constraints are effective 
to improve sampling errors. 
Section \ref{sec48site} is to report our numerical results on the $48$-site lattice  
where we know neither the exact eigenstate nor the exact eigenvalue.   
Assuming some symmetries which exist in the model on the $36$-site lattice,
we evaluate expectation values of the $m$-th power of the operator $\hat Q$ 
obtained from several approximate states whose $S_z$, 
the $z$ component of the total spin $S$ of the system, is less than or equal to $4$.    
We then estimate lowest energies for each sectors with $S_z=\kappa $ $(\kappa = 0, 1, 2, 3, 4)$.
We see our data are well described by the arguments  
based on the spontaneous symmetry breaking.  
The final section is for summary and discussions.
At the end of the paper we add three appendices in order to give detailed description 
for some parts of our numerical study. 
Appendix A is to explain how we determine values of $l$ in (\ref{qdef}).
In Appendix B we show our procedure to obtain approximate states on the 36-site and 48-site lattices. 
Finally Appendix C provides an empirical formula we use in the evaluation of the systematic 
error which is caused by employing the power method with a finite value of the power.       
  
\section{Method}
\label{sec2}

In this section we describe our method. 
Subsection \ref{subsec21} is to give a brief description of the conventional SSS method, 
the SSS method which is not accompanied with any constraints\cite{MM2}.
Then subsection \ref{subsec22} follows to show our basic idea for constraints.
Finally in subsection \ref{subsec23} we describe the constrained SSS method in detail. 

\subsection{The SSS method}
\label{subsec21}

The stochastic state selection is realized by a number of stochastic variables.
Let us expand a normalized state ${\mid \psi \rangle}$ in an $N$-dimensional 
vector space by a basis 
${\{ \mid j \rangle \}}$, ${\mid \psi \rangle} = {\sum_{j=1}^{N} \mid j \rangle c_j}$.
Then we generate a stochastic variable $\eta_j$ following to 
the {\em on-off probability function},
\begin{equation}
P_j(\eta) \equiv 
\frac{1}{a_j}\delta(\eta -a_j) +(1- \frac{1}{a_j})\delta(\eta),  
\qquad \frac{1}{a_j} \equiv \min \left( 1, \frac{|c_j|}{\epsilon} \right).
\label{probf}
\end{equation}
A positive parameter $\epsilon $ which is common to all ${P_j(\eta)}$ 
$(j=1,2,\cdots, N)$ controls the reduction rate
\footnote{If the state ${\mid \psi \rangle}$ is not normalized, $\epsilon$ in 
(\ref{probf}) should be replaced by $\epsilon \sqrt{\langle \psi \mid \psi \rangle}$}. 
Note that $\eta_j=a_j \ (\geq 1)$ or $\eta_j=0$ and
statistical averages are ${\langle \! \langle \eta_j \rangle \! \rangle} = 1$
and ${\langle \! \langle \eta_j^2 \rangle \! \rangle} = a_j$ because of (\ref{probf}).
A {\em random choice operator} $\hat{M}$ is then defined by  
\begin{equation}
\hat{M}\equiv \sum_{j=1}^N \mid j \rangle \eta_j \langle j \mid \,.
\end{equation}
Using this $\hat{M}$ we obtain a state $\mid \tilde{\psi} \rangle$,
\begin{equation} 
\mid \tilde{\psi} \rangle \equiv \hat{M} {\mid \psi \rangle }= {\sum_{j=1}^N  \mid j \rangle c_j \eta_j}, 
\end{equation}
which has less non-zero elements than ${\mid \psi \rangle}$ has.
We call the difference between $\mid \tilde{\psi} \rangle$ and $\mid \psi \rangle$,  
\begin{equation} 
\mid \chi \rangle g \equiv \mid \tilde{\psi} \rangle -\mid \psi \rangle \ ,
\label{randomstate}
\end{equation}
a {\em random state}.

Since $\langle \! \langle \eta_j \rangle \! \rangle=1$
an expectation value ${\langle \psi \mid \hat O \mid \psi \rangle }$
with an operator $\hat O$ is exactly equal to the statistical average
${\langle \! \langle \ \langle \psi \mid \hat O \hat{M} 
\mid \psi \rangle \ \rangle \! \rangle }$. 

Note that in the conventional SSS method {\em all} $\eta_j$'s 
are independently generated stochastic variables.

\subsection{Basic ideas for constraints}
\label{subsec22}
In this subsection we show our basic ideas in a very simple toy model. 
This model has only two basis states in the vector space, 
$ \mid 1 \rangle  $ and $ \mid 2 \rangle $,
which are orthonormalized as 
$\langle 1 \mid 1 \rangle = \langle 2 \mid 2 \rangle = 1$ and
$\langle  1 \mid 2 \rangle = 0$.  
If we apply the conventional SSS method to a state  
$\mid  \psi \rangle =$ $\mid 1 \rangle c_1 + \mid 2 \rangle  c_2$
$( c_1 \neq 0, \ c_2 \neq 0)$ in this vector space we obtain a state $\mid \tilde{\psi} \rangle$,
\begin{equation}
 \mid \tilde{\psi} \rangle  \equiv
 \mid 1 \rangle c_1 \eta_1 + \mid 2 \rangle c_2 \eta_2 ,
\end{equation}
where both $\eta_1$ and $\eta_2$ are stochastic variables generated by (\ref{probf}).
We can reproduce the state $\mid \psi \rangle$ by the averaging process since 
\begin{equation}
 \langle \! \langle \ \mid \tilde{\psi} \rangle \ \rangle \! \rangle =
\mid 1 \rangle c_1 \langle \! \langle\eta_1 \rangle \! \rangle
 + \mid 2 \rangle c_2 \langle \! \langle\eta_2 \rangle \! \rangle
=\mid 1 \rangle c_1 + \mid 2 \rangle c_2 =\mid \psi \rangle 
\label{psip}
\end{equation}
Now we notify that independency between $\eta_1$ and $\eta_2$ are not needed in (\ref{psip}) 
because $\langle \! \langle \ \mid \tilde{\psi} \rangle \ \rangle \! \rangle =\mid \psi \rangle$
is fulfilled as far as 
$ \langle \! \langle \eta_1 \rangle \! \rangle=\langle \! \langle \eta_2 \rangle \! \rangle=1 $.
By making use of this possible dependency we can impose a constraint. 
For example, let $\eta_2$ be {\em not} an independent stochastic variable {\em but} the following function of
$\eta_1$,
\begin{equation} 
  \eta_2= 1+ \left(\frac{c_1}{c_2}\right)^2 (1-\eta_1).
\label{eta2}
\end{equation}
It is clear that (\ref{psip}) holds with (\ref{eta2}) because
$\langle \! \langle \eta_2 \rangle \! \rangle=1 $ 
follows from $\langle \! \langle \eta_1 \rangle \! \rangle=1 $.
With (\ref{eta2}) we also see that   
\begin{equation} 
\langle \psi \mid \tilde{\psi} \rangle= c_1^2\eta_1 + c_2^2\eta_2 =c_1^2+c_2^2 = \langle \psi \mid \psi \rangle  
\label{conex}
\end{equation}
holds in each sampling. 
In other words, we have a constraint that a normalization 
$\langle \psi \mid \tilde{\psi} \rangle $ is a constant $c_1^2+c_2^2$. 
This also means that 
\begin{equation} 
\langle \psi \mid \chi \rangle g = 0 
\label{conoth}
\end{equation}
for any sampling, 
where $\mid \chi \rangle g$ is the random state defined by (\ref{randomstate}).

It is possible to impose a more general constraint 
\begin{equation} 
\langle \Phi \mid \chi \rangle g = 0   
\label{conexg}
\end{equation}
instead of (\ref{conoth}), with a given state 
$\mid \Phi \rangle 
= \mid 1 \rangle b_1 + \mid 2 \rangle b_2 \ \ \ (b_1 \neq 0, b_2 \neq 0)$.
To do so we should let 
\begin{equation} 
  \eta_2= 1+ \left(\frac{b_1 c_1}{b_2 c_2}\right) (1-\eta_1).
\label{eta2g}
\end{equation}
Then we obtain 
\begin{equation} 
\langle \Phi \mid \tilde{\psi} \rangle= b_1 c_1\eta_1 + b_2 c_2\eta_2 =
b_1 c_1+ b_2 c_2 = \langle \Phi \mid \psi \rangle ,  
\label{conexb}
\end{equation}
which is equivalent to (\ref{conexg}).

\subsection{The constrained SSS method}
\label{subsec23}
In this subsection we present a way to impose constraints in the SSS method.
It is straightforward to generalize discussions in subsection \ref{subsec22} 
for a larger vector space constructed by $N$ basis states. 
Let the constraint be, with a given state 
$\mid \Phi \rangle = \sum_{j=1}^N \mid j \rangle b_j$, 
\begin{equation}
\langle \Phi \mid \chi \rangle g =0
\label{consk1n}
\end{equation}
for a state $\mid \psi \rangle = \sum_{j=1}^N \mid j \rangle c_j$ and 
$\mid \chi \rangle g = \sum_{j=1}^N \mid j \rangle c_j (\eta_j -1)$. 
In this case we should solve the equation 
\begin{equation} 
\sum_{j=1}^N b_j c_j (\eta_j -1) =0
\label{etajeq}
\end{equation}
to give one of $\eta_j$, say $\eta_J$, as a function of $N-1$ independent stochastic variables with $j \neq J$. 
\begin{equation}
\eta_J=1+\sum_{j \neq J}\left(\frac{b_j c_j}{b_J c_J}\right)(1-\eta_j) .
\label{etajsol}
\end{equation}
From the fact that $\langle \! \langle \eta_j \rangle \! \rangle =1$ for all 
$j$ except for $J$, 
it is clear that (\ref{etajsol}) guarantees 
$\langle \! \langle \eta_J \rangle \! \rangle =1$.  
As for $\langle \! \langle \eta_J^2 \rangle \! \rangle $, we find  
\begin{equation}
 \langle \! \langle \eta_J^2 \rangle \! \rangle - 1 =
 \sum_{j \neq J} \left( \frac{b_j c_j}{b_J c_J} \right) ^2 (a_j-1).
\label{etaj2g} 
\end{equation}
In principle, $J$ can be any of $1$, $2$, $\cdots$, $N$. 
From practical point of view, however, $J$ should be chosen carefully 
so that sampling errors are diminished.
When we choose the state $\mid \psi \rangle$ as the state $\mid \Phi \rangle$, 
which means $b_j=c_j$ for all $j$, the right-hand side of 
(\ref{etaj2g}) becomes $\sum_{j \neq J} \left( \frac{c_j}{c_J} \right) ^4 (a_j-1)$.
In this case it is clear that we can lessen this quantity by picking up $J$ 
which realizes $|c_J| \geq |c_j|$ for any $j$.
Numerical examinations for this choice will be given in following sections.

Another generalization to impose more than one constraint is also easy. 
Let $K$ denote the number of constraints. As far as $K < N$ we can impose constraints   
for given states $\mid \Phi_k  \rangle \ (k=1,2,\cdots,K)$  
\begin{equation}
 \langle \Phi_k \mid \chi \rangle g = 0,  \ \ \ ( k=1,2,\cdots ,K) ,
\label{consk}
\end{equation}
by requesting $K$ variables among $\eta_j$'s depend on $N-K$ other $\eta_j$'s which are 
independent stochastic variables generated by (\ref{probf}). 
More concrete description of the way to impose constraints is as follows.
If we know the coefficients of the expansions of these states,
the above constraints (\ref{consk}) read 
\begin{equation}
  \sum_j b_j^{(k)} c_j(\eta_j-1)= 0 \ \ \ ( k=1,2,\cdots ,K) ,
\label{conskeq}
\end{equation}
with
\begin{equation}
 \mid \psi \rangle = \sum_{j=1}^N \mid j \rangle c_j , \ \ \
 \mid \Phi_k \rangle = \sum_{j=1}^N \mid j \rangle b_j^{(k)} .
\end{equation}
From (\ref{conskeq}) we obtain 
$\eta_{J_k},(k=1,..,K)$ which are dependent on other stochastic variables.
Since these dependencies are linear,
the conditions $\langle \! \langle \eta_{J_k} \rangle \! \rangle =1 $
are always satisfied.
A choice of $\{ J_k \}$ is arbitrary, but the sample fluctuations will depend on the choice.

Hereafter we denote a random choice operator used in the constrained SSS method 
by $\hat M_{\rm c} $ in order to avoid confusions with the random choice operator 
in the conventional SSS method. 

\section{Study with an exact eigenstate}
\label{secexact}

In this section we study a system for which we know an exact energy eigenstate
as well as its eigenvalue. 
First we make analytical discussions with the exact eigenstate 
$\mid \psi_{\rm E} \rangle$ whose exact eigenvalue is $E$. 
Then we numerically study the spin 1/2 quantum Heisenberg antiferromagnet 
on a 16-site triangular lattice as a concrete example. 
For this small lattice we can easily obtain the exact ground state 
and its eigenvalue by the exact diagonalization. 
This enables us to compare our results from the constrained SSS method with 
the exact ones.

Let us examine the expectation value of the $m$-th power of $\hat Q = l\hat{I}-\hat{H}$, 
$\langle \psi \mid \hat Q ^ m \mid \psi \rangle $, with  
$\mid \psi \rangle = \mid \psi_{\rm E} \rangle $.
From the exact eigenvalue we obtain 
\begin{equation}
\langle \psi_{\rm E} \mid \hat{Q}^m \mid \psi_{\rm E} \rangle 
= Q^m, \ \ \ Q \equiv l-E . 
\end{equation}
In order to calculate these expectation values by the constrained SSS method,
we insert random choice operators $\hat M _{\rm c}^{(n)} \ (n=1,2,\cdots,m)$ and calculate 
\begin{equation}
\langle \psi_{\rm E} \mid \hat{Q}\hat{M}_{\rm c}^{(m)} \cdots \hat{Q}\hat{M}_{\rm c}^{(1)} \mid 
\psi_{\rm E} \rangle .
\label{sssexv}
\end{equation}
Here we denote random choice operators by $\hat{M}_{\rm c}^{(n)} $ 
instead of $\hat M_{{\rm c}} $ since we want to emphasize that different operators, 
each of which includes stochastic variables independent of those in other operators, 
are used.
Let us define states $\mid \phi^{(n)} \rangle$ as follows,
\begin{equation}
\mid \phi^{(n)} \rangle \equiv \hat Q \hat 
M _{\rm c} ^{(n-1)} \cdots  \hat Q \hat M _{\rm c}^{(1)} \mid \psi_{\rm E} \rangle 
\ \ \ ( n \geq 2), \ \ \ \mid \phi^{(1)} \rangle \equiv \mid \psi_{\rm E} \rangle.
\end{equation}
Note that  
$\mid \phi^{(n)} \rangle = \hat{Q}\hat{M}_{\rm c} ^{(n-1)} \mid \phi^{(n-1)} \rangle $  
with $n \geq 2$ by definition.
For each $\hat{M}_{\rm c}^{(n)} $ we impose the dependency (\ref{etajsol}) 
so that the constraint 
\begin{equation}
\langle \psi_{\rm E} \mid \chi^{(n)} \rangle g^{(n)}= 0
\label{conson}
\end{equation}
holds for 
\begin{equation}
\mid \chi^{(n)} \rangle g^{(n)} \equiv \hat M_{\rm c}^{(n)} \mid \phi^{(n)} \rangle - 
\mid \phi^{(n)} \rangle . 
\end{equation}
Here $J^{(n)}$ of the dependent stochastic variable is determined by the condition 
$|c^{(n)}_{J^{(n)}}|=\max_{1\leq j \leq N}|c^{(n)}_j|$ with the expansion 
$\mid \phi^{(n)} \rangle = \sum \mid j \rangle c^{(n)}_j$.  
Notifying that 
$\langle \psi_{\rm E} \mid \hat{M}_{\rm c}^{(n)}  \mid \phi^{(n)} \rangle 
=\langle \psi_{\rm E} \mid  \phi^{(n)} \rangle 
$
follows from (\ref{conson}), we obtain
\begin{eqnarray}
\nonumber
\langle \psi_{\rm E} \mid \hat{Q}\hat{M}_{\rm c}^{(n)} \mid \phi _n\rangle  & =& 
Q \langle \psi_{\rm E} \mid \hat{M}_{\rm c}^{(n)} \mid \phi^{(n)} \rangle 
= Q\langle \psi_{\rm E} \mid  \phi^{(n)} \rangle \\ 
&=& Q \langle \psi_{\rm E} \mid \hat{Q}\hat{M}_{\rm c}^{(n-1)} \mid \phi^{(n-1)} \rangle .
\label{reduc}
\end{eqnarray}
Using (\ref{reduc}) repeatedly, we find 
\begin{eqnarray}
\nonumber
\langle \psi_{\rm E} \mid \hat{Q}\hat{M}_{\rm c}^{(m)} \cdots \hat{Q}\hat{M}_{\rm c}^{(1)}
\mid \psi_{\rm E} \rangle  
&=& \langle \psi_{\rm E} \mid \hat Q \hat{M}_{\rm c}^{(m)} \mid \phi^{(m)} \rangle  
\\ \nonumber 
&=& Q \langle \psi_{\rm E} \mid \hat{Q} \hat M_{\rm c} ^{(m-1)} \mid \phi^{(m-1)} \rangle 
\\ \nonumber
& = & \cdots
\\ \nonumber 
& = & Q^{m-1} \langle \psi_{\rm E} \mid \hat{Q} \hat M_{\rm c}^{(1)} \mid \phi^{(1)} \rangle      
\\ \nonumber 
&=& Q^m \langle \psi_{\rm E} \mid \hat M_{\rm c}^{(1)} \mid \phi^{(1)} \rangle 
=Q^m \langle \psi_{\rm E} \mid \phi^{(1)} \rangle 
\\ 
&=& Q^m \langle \psi_{\rm E} \mid \psi_{\rm E} \rangle = Q^m.
\end{eqnarray}
This result implies that the exact calculation is possible in the sampling. 
Note that only one sampling is enough in the constrained SSS method.

\begin{figure}[ht]
\begin{center}
\scalebox{0.45}{\includegraphics{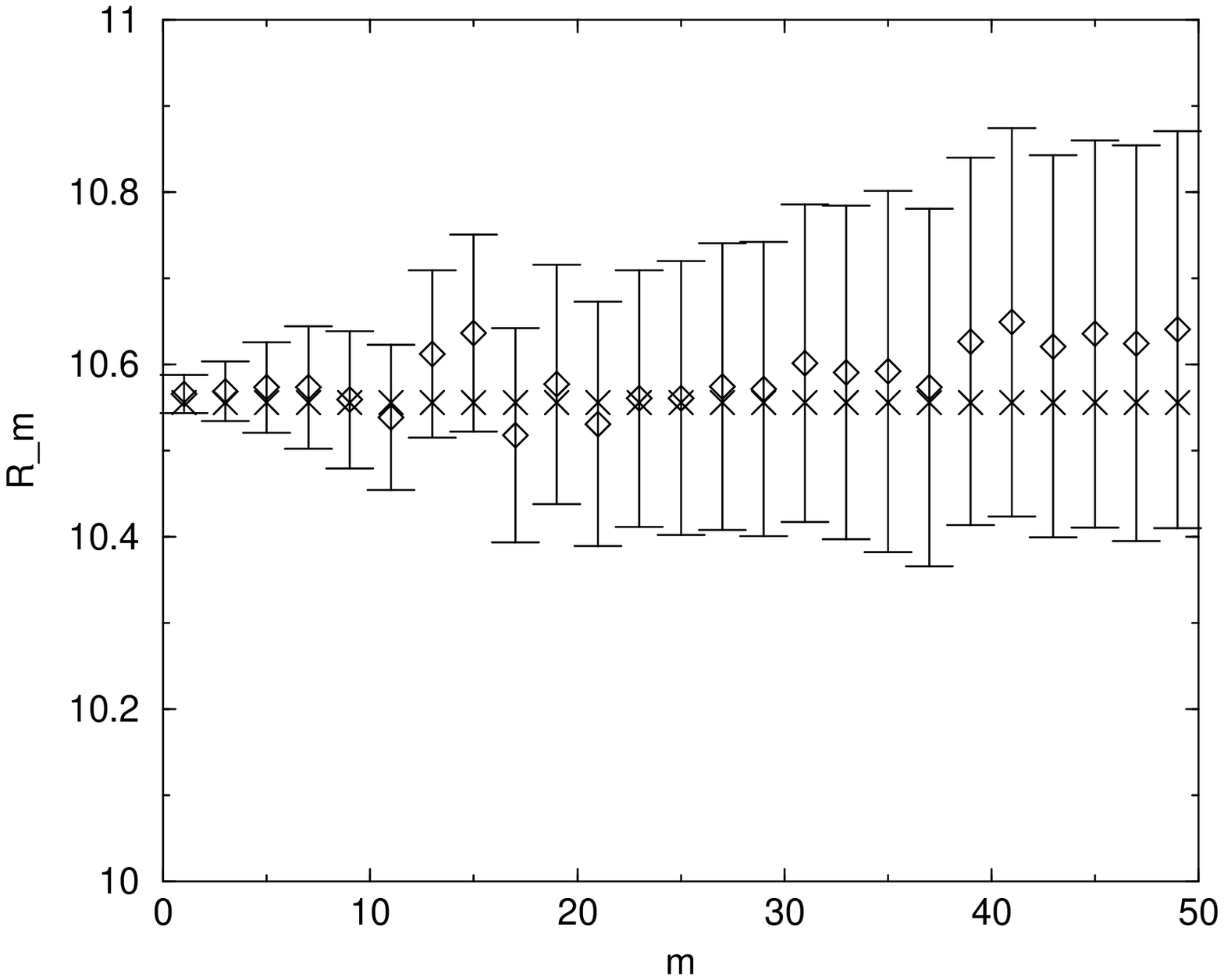}}
\caption{Ratios defined by (\ref{ratiom}) calculated on a 
16-site triangular lattice with $\epsilon = 0.05$. 
Here we use the exact ground state as $\mid \psi_{\rm E} \rangle$ for which 
$E=E_{\rm g} = -8.5555$.
The crosses plot data from one sample obtained by the constrained SSS method. 
We also plot, by open diamonds, ratios
$\langle \psi_{\rm E} \mid 
\hat{Q}\hat{M}_{m} \cdots \hat{Q}\hat{M}_1\mid \psi_{\rm E} \rangle 
/\langle \psi_{\rm E} \mid \hat{Q}\hat{M}_{m-1} \cdots 
\hat{Q}\hat{M}_1\mid \psi_{\rm E} \rangle $ 
which are averages from 100 samples generated 
by the conventional SSS method.
Errors in the figure are statistical errors only.
}
\label{fig:16site_a}
\end{center}
\end{figure}

Now we present numerical results for the ground state of the system on the 16-site 
triangular lattice. The ground state energy is known to be $E=E_{\rm g}=-8.5555$.
Since we employ $l=2$ here as is stated in Appendix A, 
the exact value of $Q$ is $l-E_{\rm g} =10.5555$. 
In Fig.~\ref{fig:16site_a} we plot ratios of the expectation values (\ref{sssexv}),
\begin{equation}
R^{(m)} \equiv  \frac{ \langle \psi_{\rm E} \mid \phi^{(m+1)} \rangle}
{ \langle \psi_{\rm E} \mid \phi^{(m)} \rangle} 
=\frac{ \langle \psi_{\rm E} \mid \hat Q  \hat{M}_{\rm c}^{(m)} \mid \phi^{(m)} \rangle}
{ \langle \psi_{\rm E} \mid \phi^{(m)} \rangle}  ,
\label{ratiom}
\end{equation}
obtained in one sampling with the constrained SSS method.
The parameter $\epsilon$ in (\ref{probf}) is $0.05$.
As is expected from above discussion, we observe $R^{(m)}=Q$ for any value of $m$.
We also present results obtained by the conventional SSS method, 
averages of $
\langle \psi_{\rm E} \mid \hat{Q}\hat{M}^{(m)}\cdots \hat{Q}\hat{M}^{(1)}\mid 
\psi_{\rm E} \rangle  /\langle \psi_{\rm E} \mid \hat{Q}\hat{M}^{(m-1)} \cdots 
\hat{Q}\hat{M}^{(1)}\mid \psi_{\rm E} \rangle $ 
from 100 samples with $\epsilon=0.05$, where $\hat M^{(n)}$'s denote
different random choice operators in the conventional SSS method.
Statistical errors for these averages are also plotted in the figure,    
where for data $X_i$ $(i=1,2, \cdots, n_{\rm sample})$ 
the statistical error $\Delta [X]$ is defined by 
\begin{equation}
\Delta [X] = \frac{\sigma [X]}{\sqrt{n_{\rm sample} -1 }},
\label{err} 
\end{equation}
with the standard deviation 
\begin{equation} 
\sigma [X]= \sqrt{ \frac {1} {n_{\rm sample}} \sum_{i=1}^{n_{\rm sample}} X_i ^2    
- \Bigl\{ \frac {1} {n_{\rm sample}} \sum_{i=1}^{n_{\rm sample}} X_i \Bigr\} ^2 }.
\label{sdev}
\end{equation}
Comparing these data we clearly see 
that fluctuations existing in the conventional SSS method  
disappear with constraints stated by (\ref{conson}).

\section{Study with an approximate eigenstate}
\label{secappx}
 
In this section we argue the case in which we use an approximate eigenstate 
$\mid \psi _{\rm A} \rangle$ instead of $\mid \psi _{\rm E} \rangle$.
First we present an analytical argument which 
endorses that the constrained SSS method is effective for eigenvalue evaluations
starting from an approximate state. 
Then using the 36-site lattice we demonstrate that, starting with an approximate ground state, 
we can obtain in the constrained SSS method the correct ground state energy with much less 
fluctuations than those from the conventional SSS method. 

Let us start with an approximate state $\mid \psi_{\rm A} \rangle$ 
which has some overlap with the corresponding exact eigenstate $\mid \psi_{\rm E} \rangle $,
\begin{equation}
\mid \psi_{\rm A} \rangle  = w\mid \psi_{\rm E} \rangle + s\mid \zeta \rangle . 
\label{psia}
\end{equation}
Here $\mid \psi_{\rm A} \rangle $ is normalized with $w^2+s^2=1$ 
and we expect $|w| \gg |s|$. 
Instead of (\ref{sssexv}) we calculate the expectation value 
$\langle \psi_{\rm A} \mid \hat{Q}\hat{M}_{\rm c}^{(m)} 
\cdots \hat{Q}\hat{M}_{\rm c}^{(1)}  \mid \psi_{\rm A} \rangle $ 
with constraints 
\begin{equation}
\langle \psi_{\rm A} \mid  \hat M _{\rm c}^{(n)} \mid \phi_{\rm A}^{(n)} \rangle =
\langle \psi_{\rm A} \mid \phi_{\rm A}^{(n)} \rangle \ \ \ (n=1,2,\cdots,m)  
\label{cphian}
\end{equation}
where 
\begin{equation}
\mid \phi_{\rm A}^{(n)} \rangle \equiv \hat Q \hat M _{\rm c} ^{(n-1)} \cdots  \hat Q 
\hat M _{\rm c}^{(1)} \mid \psi_{\rm A} \rangle 
\ \ \ ( n \geq 2), \ \ \ \mid \phi_{\rm A}^{(1)} \rangle \equiv \mid \psi_{\rm A} \rangle.
\label{dphian}
\end{equation}
Let us examine $\langle \psi_{\rm A} \mid \hat Q \hat M_{\rm c}^{(1)} \mid 
\phi_{\rm A}^{(1)} \rangle$ 
when the constraint (\ref{cphian}) holds. Using (\ref{psia}) we find that
\begin{eqnarray}
\nonumber
\langle \psi_{\rm A} \mid \hat Q \hat M_{\rm c}^{(1)} \mid \phi_{\rm A}^{(1)} \rangle  
&=& \Bigl( w\langle \psi_{\rm E} \mid + s\langle \zeta \mid \Bigr) \hat Q 
\hat M_{\rm c}^{(1)} \mid \phi_{\rm A}^{(1)} \rangle \\ \nonumber 
&=& \left( Qw\langle \psi_{\rm E} \mid + s\langle \zeta \mid \hat Q \right) 
\hat M_{\rm c}^{(1)} \mid \phi_{\rm A}^{(n)} \rangle \\ \nonumber  
&=& \left( Q\bigl[ \ \langle \psi_{\rm A} \mid -s\langle \zeta \mid \ \bigr] 
+ s\langle \zeta \mid \hat Q \right) \hat M_{\rm c}^{(1)} \mid  \phi_{\rm A}^{(1)} \rangle \\ \nonumber
&=& Q\langle \psi_{\rm A} \mid \hat M_{\rm c}^{(1)} \mid \phi_{\rm A}^{(1)} \rangle + s  
\langle \zeta \mid \left(\hat Q -Q\hat I \right)\hat M_{\rm c}^{(1)} \mid \phi_{\rm A}^{(1)} \rangle
\\ \nonumber
&=& Q\langle \psi_{\rm A} \mid \phi_{\rm A}^{(1)} \rangle 
+ s \langle \zeta \mid \left(\hat Q -Q\hat I \right) 
\Bigl(\mid \phi_{\rm A}^{(1)}\rangle + \mid \chi_{\rm A}^{(1)} \rangle g_{\rm A}^{(1)} \Bigr) 
\\ \nonumber
&=& Q\langle \psi_{\rm A} \mid \psi_{\rm A} \rangle 
+ s \langle \zeta \mid \left(\hat Q -Q\hat I \right) 
\Bigl(\mid \psi_{\rm A} \rangle + \mid \chi_{\rm A}^{(1)} \rangle g_{\rm A}^{(1)} \Bigr) \\ \nonumber 
&=& Q + s \langle \zeta \mid \left(\hat Q -Q\hat I \right) \mid \psi_{\rm A} \rangle \\ 
&+& s \langle \zeta \mid \left(\hat Q -Q\hat I \right) \mid \chi_{\rm A}^{(1)} 
\rangle g_{\rm A}^{(1)}, 
\label{zetaaqm1}
\end{eqnarray}
where 
\begin{equation}
\mid \chi_{\rm A}^{(n)} \rangle g_{\rm A}^{(n)} \equiv 
\hat M_{\rm c}^{(n)} \mid \phi_{\rm A}^{(n)} \rangle -  \mid \phi_{\rm A}^{(n)} \rangle .  
\label{chingna}
\end{equation}
Note that in the right-hand side of (\ref{zetaaqm1})
only the last term 
contains the fluctuation by a sampling and that this term should be small when 
$|s| \ll 1$.

Next we turn to the numerical study for the ground state on the 36-site lattice.
The exact ground state energy of this system is known to be $-0.186791$ per bond\cite{Bernu}, namely $-20.1734$ in total. 
Under the symmetries the ground state of this system has, 
the number of basis states in the whole $S_z=0$ sector amounts to $\sim 2.2 \times 10^7$.
Following the procedures described in Appendix B
we create two $\mid \psi _{\rm A} \rangle$'s for the ground state of the system. 
Let us denote them by $\mid \psi_{{\rm A}\mu} \rangle$ $(\mu=1,2)$.
The number of basis states with non-zero coefficients in the expansion of  
$\mid \psi_{{\rm A}1} \rangle$ is $333001$ and the expectation value of
$\hat H$ is $\langle \psi_{{\rm A}1} \mid \hat H \mid \psi_{{\rm A}1} \rangle = -19.577$,
while $\mid \psi_{{\rm A}2}\rangle$ includes $887875$ basis states with non-zero coefficients
and $\langle \psi_{{\rm A}2} \mid \hat H \mid \psi_{{\rm A}2} \rangle = -19.817$.
In the same manner as was stated in section \ref{secexact}, we request one constraint 
for each random choice operator. We choose $J^{(n)}$ for the only non-independent variable $\eta_{J^{(n)}}$ using the same criteria as that in the $N_s=16$ case.
Based on conditions we notify in Appendix A, we choose $l=4$ 
(namely $\hat Q = 4 \hat I - \hat H$) here.
Our results for the system are presented in Figs. \ref{fig:36site_a}, \ref{fig:36site_b}, \ref{fig:36site_c} and \ref{fig:36site_d}. 
Fig.~\ref{fig:36site_a} plots  
\begin{equation}
R_{\rm A}^{(m)} \equiv  \frac{ \langle \psi_{\rm A} \mid \phi_{\rm A}^{(m+1)}  \rangle}
{ \langle \psi_{\rm A} \mid \phi_{\rm A}^{(m)} \rangle} 
=\frac{ \langle \psi_{\rm A} \mid \hat Q  \hat{M}_{\rm c}^{(m)} \mid \phi_{\rm A}^{(m)} \rangle}
{ \langle \psi_{\rm A} \mid \phi_{\rm A}^{(m)} \rangle}  
\label{ratioam}
\end{equation}
calculated from $100$ samples with the constraint stated by (\ref{cphian}). 
The value of the parameter $\epsilon $ is $0.01$. 
In the figure we also plot values of 
$\langle \psi_{\rm A} \mid \hat{Q}\hat{M}^{(m)} \cdots 
\hat{Q}\hat{M}^{(1)}\mid \psi_{\rm A} \rangle  /
\langle \psi_{\rm A} \mid \hat{Q}\hat{M}^{(m-1)} \cdots 
\hat{Q}\hat{M}^{(1)}\mid \psi_{\rm A} \rangle $, 
which are obtained by the conventional SSS method from averages of  
$100$ samples with $\epsilon = 0.01$ using the approximate state 
$\mid \psi_{{\rm A}1} \rangle$.  
We see that the values mostly agree in both method.
In Fig.~\ref{fig:36site_b} we compare the standard deviations $\sigma$ of the ratios 
shown in Fig.~\ref{fig:36site_a}.
We see the exponential growth of the standard deviations except for the $m=1$ datum 
from the constrained SSS method.  
It is quite impressive that fluctuations from the constrained SSS method are much less than 
those from the conventional SSS method.

Let us present some more data on the 36-site lattice 
which will be helpful to understand the role of the parameter 
$\epsilon$ and the quality of the approximate state $\mid \psi_{\rm A} \rangle $. 
Fig.~\ref{fig:36site_c} is to show how much basis states are included in the expansion of  
$\hat M _{c}^{(m)} \mid \phi _{\rm A}^{(m)} \rangle$.
Here we present only results from $\mid \psi _{{\rm A}1} \rangle$ 
since we observe $N_{\rm a}$'s are mostly insensible to the choice of the 
$\mid \phi_{\rm A}^{(1)} \rangle = \mid \psi _{\rm A} \rangle$ stated above. 
We see that as $m$ increases $N_{\rm a}$ becomes almost constant for each value of $\epsilon$ 
and there 
\begin{equation}
N_{\rm a} \propto \epsilon ^{-1}
\label{navseps}
\end{equation}
holds. Note that this means that by the choice of $\epsilon$  
we can change the CPU time and the memory which we should supply in numerical studies.    
Fig.~\ref{fig:36site_d} shows several values of $R_{\rm A}^{(20)}$ from 10 samples 
obtained with different values of $\epsilon$ as well as with 
two different approximate states $\mid \psi _{{\rm A}1} \rangle$ and 
$\mid \psi _{{\rm A}2} \rangle$.
It is quite reasonable that the data indicate we can obtain a better lower bound for $Q$
(a better upper bound for $E$) with a better approximate state 
$\mid \psi_{{\rm A}2} \rangle$ when the value of $m_{\rm max}$ is the same.
In order to prepare a better approximate state, however, 
we of course have to deal with a larger portion of the Hilbert space. 
We also see in Fig.~\ref{fig:36site_d} that statistical errors, and therefore the standard deviation 
which is $3 \Delta [R_{\rm A}^{(20)}]$ in this figure with $\sqrt{n_{\rm sample} -1}=3$,
are irrelevant to the choice of the approximate state, while they  
decrease rapidly for less value of $\epsilon$. 
We observe that  
\begin{equation}
\sigma [R_{\rm A}^{(20)}] \propto \epsilon ^ \gamma  \ \ \ (\gamma \sim 3) . 
\label{sig20vseps}
\end{equation}

\section{Application to the 48-site system}
\label{sec48site}

\begin{figure}[h]
\begin{center}
\scalebox{0.45}{\includegraphics{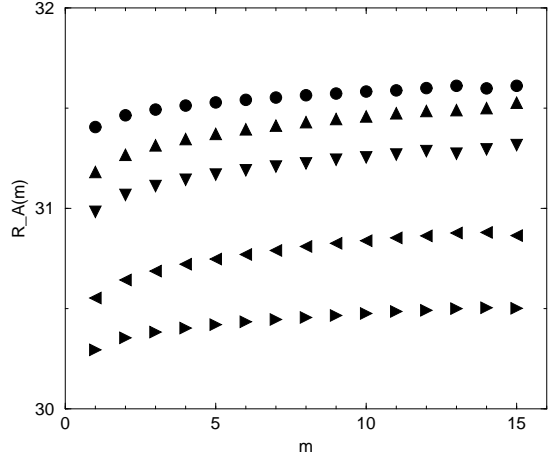}}
\caption{
Ratios (\ref{ratioam}) in the $S_z=\kappa$ ($0 \leq \kappa \leq 4$) sectors 
on a 48-site triangular lattice calculated from 100 samples    
with $\epsilon=5 \times 10^{-3}$ ($\epsilon = 7 \times 10^{-3}$) for $S_z=0$ ($S_z \geq 1$).
We use approximate states which are composed of $489413140$, 
$150733425$, $115759910$, $73294432$ and $90008649$ basis states 
in the $S_z=0,1,2,3$ and 4 sectors, respectively.
Circles present the $S_z=0$ results, while   
triangles-up, -down, -left and -right show results for $S_z=1,2,3$ and 4. 
All statistical errors are within the marks.
}
\label{fig:48site_a}
\end{center}
\end{figure}

In this section we study the ground state energy and some excited energies of the spin  
1/2 quantum Heisenberg antiferromagnet 
on the $48$-site triangular lattice using the constrained SSS method.
Similarly in sections \ref{secexact} and \ref{secappx}, we request one constraint 
for each random choice operator and decide $J^{(n)}$ for the only non-independent variable $\eta_{J^{(n)}}$ with the condition $|c^{(n)}_{J^{(n)}}| \geq |c^{(n)}_j|$ for all $j$. 
In each sector with $S_z=\kappa \ (\kappa=0,1,2,3,4)$ we study the state 
whose energy eigenvalue is the lowest in that sector.    
Following the procedures in Appendix B 
we calculate $\mid \psi_{\rm A} \rangle$'s summarized in Table~1.
As is stated in Appendix B, 
symmetries we assume for the approximate states are the same as those for the $N_s=36$ case. 
Figure~\ref{fig:48site_a} shows our results for 
the ratio $R_{\rm A}^{(m)}$ defined by (\ref{ratioam}) up to $m_{\rm max}=15$
with values of $l$ given in Appendix A.
All data are calculated from 100 samples with $\epsilon = 5 \times 10^{-3}$ 
($\epsilon = 7 \times 10^{-3}$ ) for $S_z =0$ ($S_z > 0$).
  
It is known that the argument by the spontaneous symmetry breaking in semi-classical
N\'{e}el ordered antiferromagnets suggests the following energy spectrum
on finite-sized lattices\cite{rev,Trumper,Neuberger,Bernu2},
\begin{equation}
 E(S)-E(0) = \frac{1}{2\chi_\triangle} \cdot \frac{S(S+1)}{N_s},
\label{ssb}
\end{equation}
where $E(S)$ and $\chi_\triangle$ denote 
the lowest energy of the system with the total spin $S$ and the susceptibility,
as long as $S \ll \sqrt{N_s}$.
Keeping this in mind, we plot values for $\overline{E}_{\kappa}/(3N_s)$ 
versus $S(S+1) $ in Fig.~\ref{fig:48site_b} with an assumption $S=\kappa$, where 
$\overline{E}_{\kappa}$ denotes the upper bound of $E$ with $S_z=\kappa$ given by 
$R_{\rm A}^{(m_{\rm max})}$. 
Namely, $\overline{E}_{\kappa}$ is $l -R_{\rm A}^{(15)}$ in the $S_z=\kappa$ sector.
Values of $\overline{E}_{\kappa}$ are also presented in Table~1.      
We see that those energies on the 48-site lattice are well described by (\ref{ssb}). 
From the least square fit of the data with $S=\kappa=0,1,2,3 $ and 4 we obtain 
$1/2\chi_\triangle=5.6$.
In Fig.\ref{fig:48site_b} we also plot the data on the 36-site lattice obtained 
by the exact diagonalization\cite{Bernu},
which gives $1/2\chi_\triangle = 5.1$. 
The fact that this value is compatible with the discussion for the susceptibility 
in ref.~\cite{Trumper} supports the finite size arguments (\ref{ssb}) 
based on the symmetry breaking. 

Finally let us comment on the spin gap $\Delta_{\rm spin} (N_s)$. 
Results on $\overline{E}_0$ and $\overline{E}_1$ give us 
$\Delta_{\rm spin} (48) = 0.284 \pm 0.027$.
Through the finite-size extrapolation using data $\Delta_{\rm spin} (12)$ 
$\Delta_{\rm spin} (36)$ and $\Delta_{\rm spin} (48)$ we obtain 
$\Delta_{\rm spin} (\infty) \sim 0.10$, which is smaller than the one 
evaluated from the data with $N_s \leq 36$\cite{rev}.  

\vskip 1cm 

\begin{table*}[h]
\begin{center}
\begin{tabular}{|c|c|c|c|c|c|} \hline
$S_z $ & $\cal{N}_{\rm A}$ & $\cal{N}_{\rm A}$/$N_\kappa$  &
$\langle \psi_{\rm A} \mid \hat H \mid \psi_{\rm A} \rangle$ 
& $\epsilon$ & $\overline{E}_\kappa$  \\ \hline
0  & $4.9 \times 10^8$ &
 $8.8 \times 10^{-3}$ & $-26.411$  &  $5 \times 10^{-3} $ & $ -26.611 \pm 0.015$ \\ \hline
1 &$1.5 \times 10^8$& 
$2.8 \times 10^{-3}$ &  $-25.981$  &  $7 \times 10^{-3} $ & $ -26.327 \pm 0.023$ \\ \hline
2 &$1.2 \times 10^8$& 
$2.4\times 10^{-3}$ & $-25.482$  &  $7 \times 10^{-3} $ & $ -25.815 \pm 0.020$ \\ \hline
3  & $7.3 \times 10^7$ & 
$1.9 \times 10^{-3}$ & $-24.753$  &  $7 \times 10^{-3} $ & $ -25.064 \pm 0.017$ \\ \hline
4  &$9.0 \times 10^7$& 
$3.1 \times 10^{-3}$ & $-24.094$  &  $7 \times 10^{-3} $ & $ -24.302 \pm 0.016$ \\ \hline
\end{tabular}
\caption{Approximate states employed for the 48-site system and results 
for the ratio $R_{\rm A}^{(m_{\rm max})}$ with $m_{\rm max}=15$,  
which we obtain from 100 samples using the constraint (\ref{cphian}).
$N_\kappa$ denotes the number of basis states of the whole $S_z=\kappa$ sector 
with the assumed symmetries, while  
$\cal{N}_{\rm A}$ is the number of basis states whose coefficients are non-zero
in the expansion of $\mid \psi_{\rm A} \rangle $. 
} 
\end{center}
\end{table*} 

\newpage

\section{Summary and discussions}
\label{secsumdis}

In previous works\cite{MM1,MMss,MM2,MM3,MMtri,MMeq} 
we have developed the stochastic state selection (SSS) method,
which is a kind of Monte Carlo method suitable to calculate eigenvalues in large quantum 
systems. In this paper we proposed 
the constrained stochastic state selection method, where some constraints are 
imposed in each sampling.   

It is a characteristic feature of the SSS method to sample many states simultaneously. 
Namely, in the SSS method we introduce $N$ stochastic 
variables to form a random choice operator, 
where $N$ denotes the number of the basis states of the whole vector space.  
In the conventional SSS method all of these $N$ stochastic variables are independent 
of each other.
In the constrained SSS method, on the other hand, $K$ variables in the random choice operator are determined by values of $N-K$ other ones which are independently generated stochastic variables.
Using these dependencies we can force $K$ relations to represent constraints
should hold in each sampling, provided that these relations are linear of all 
stochastic variables.

We pursue arguments to a conclusion that  
the constrained SSS method is effective to decrease statistical errors in calculations 
of energy eigenvalues using approximate eigenstates.
Numerical demonstrations then follow, which apply the constrained SSS method to 
the spin-1/2 quantum Heisenberg antiferromagnet
on a 36-site triangular lattice. 
Here we 
employ the power method in combination with the constrained SSS samplings.
We impose one constraint which requests that each sample for a given state should not change the inner product between the initial approximate state and the given state.
With initial states which approximate the ground state of the system, we calculate
expectation values to be used in the power method.
We observe much less fluctuations in the constrained SSS method 
compared to those in the conventional SSS method and  
the ground state energy estimated in the constrained SSS method are in good agreement 
with the known exact value.
   
We then successfully calculate the lowest energies in $S_z=\kappa$ ($\kappa=0,1,2,3,4$) sectors on the 48-site triangular lattice.  
Our results on low-lying energies with different values of $S_z$  
add an evidence to the ordered ground state and
the finite-size arguments based on the symmetry breaking.
Especially, our result on the spin susceptibility obtained from the ground state energy 
and the low-lying excited state energies 
is consistent with finite-size effects reported in ref.~\cite{Trumper}. 

Several comments are in order.

In study of the $S_z=0$ sector in ref.~\cite{MMtri} 
we use the Lanczos method together with the conventional SSS method. 
In the present work we employ a simple power method instead of the Lanczos method so that
the calculation algorithm is simple enough to enable us to calculate 
eigenvalues with $S_z \neq 0$ within our computer resources. 

\begin{figure}[h]
\begin{center}
\scalebox{0.45}{\includegraphics{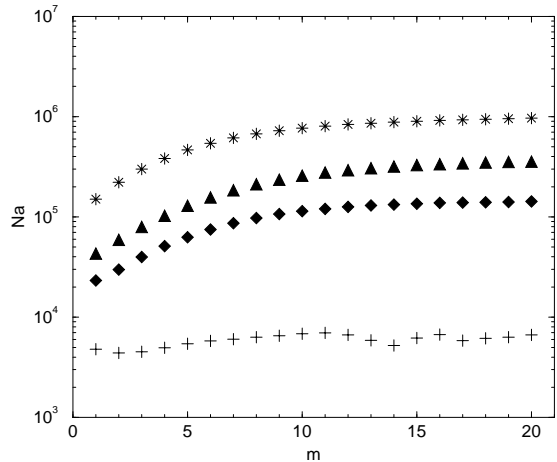}}
\caption{$N_a$, the number of basis states with non-zero coefficients 
in the expansion of $\hat M_{\rm c}^{(m)} \mid \phi_{\rm A}^{(m)} \rangle$.
All data are calculated from one sample with 
$\mid \psi_{\rm A} \rangle = \mid \psi_{{\rm A}1} \rangle$.
Pluses, diamonds, triangles and asterisks present the results with 
$\epsilon = 5 \times 10^{-2}$, $1 \times 10^{-2}$, $5 \times 10^{-3}$ and 
$1 \times 10^{-3}$, respectively.
}
\label{fig:36site_c}
\end{center}
\end{figure}

The efficiency of the method in numerical studies 
is mainly controlled by the value of the parameter $\epsilon$.
In general we can expect better results with smaller values of $\epsilon$,
but those calculation would take more CPU time and memory.   
Therefore in actual calculations 
one should choose the value of $\epsilon$ so that the  
computer resources stay within limits of his computers\footnote{
Note that the method is applicable even with large values of $\epsilon$ provided that 
the number of samples is large enough.}.  
Roughly speaking, the memory size $M_{\rm CPU}$ is proportional to
$N_{\rm b}$, the number of basis states which have 
non-zero coefficients in the expansion of $\mid \phi ^{(m+1)} \rangle = \hat Q
\hat M _{\rm c}^{(m)} \mid \phi ^{(m)} \rangle$. 
Since we observe that $N_{\rm b}$ is proportional to $N_{\rm a}$ shown in 
Fig.~\ref{fig:36site_c} and 
that its dependency on $\epsilon$ is described as (\ref{navseps}), it leads 
\begin{equation}
M_{\rm CPU} \ \propto \ \epsilon ^{-1}.  
\label{cpumem}
\end{equation}
The total CPU time $T_{\rm CPU}$, on the other hand, has a more complicated relation 
with $\epsilon$ because it depends on the number of samples 
$n_{\rm sample}$ as well as $N_{\rm a}$. It also depends on 
the number of iterations $m_{\rm max}$ when we use the method in combination with 
the power method. Therefore 
\begin{equation}
T_{\rm CPU} \ \propto \ m_{\rm max}  \cdot n_{\rm sample} \cdot N_{\rm a}.  
\label{cputim}
\end{equation}
Values of $m_{\rm max}$ and $n_{\rm sample}$ are determined from the results for 
the standard deviations $\sigma$ such as those shown in Fig.~\ref{fig:36site_b}.
Since, as we have observed an exponential growth of $\sigma$ as a function of $m$ 
(except for $m=1$) there, fluctuations of the data grow rapidly when iterations are
repeated. We therefore have to give up our numerical study with some finite 
value of $m$ before the data become statistically meaningless. If we can employ a smaller value of $\epsilon $, the value of $m_{\rm max}$ would become larger 
because of much improvement of $\sigma$ suggested by (\ref{sig20vseps}).  
The number of samples should be chosen so that the statistical
errors are reasonably small for $m \leq m_{\rm max}$. 
The total CPU time we spent to calculate the data presented in this paper is 
about $2000$ hours
with a computer whose memory is 8 Giga Bytes and whose CPU is Xeon Dual Core.

\begin{figure}[h]
\begin{center}
\scalebox{0.45}{\includegraphics{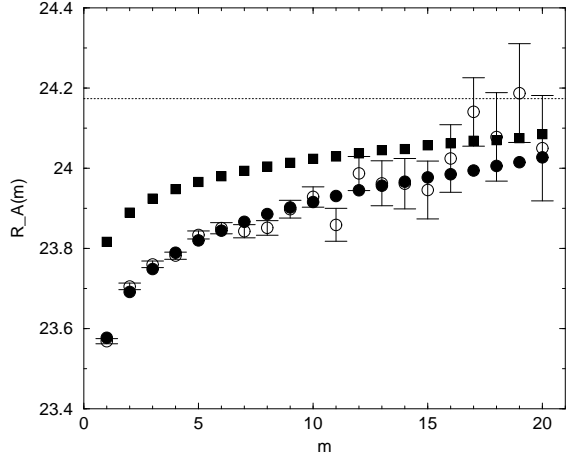}}
\caption{Ratios (\ref{ratioam}) on a 36-site triangular lattice 
which are generated by the constrained SSS method (filled marks).
We employ two states to approximate the ground state energy.
By filled circles we plot results which we obtain 
using an approximate state constructed by $333001$ basis states 
($\mid \psi_{{\rm A}1} \rangle$).
With the same approximate state we also calculate ratios
$ \langle \psi_{\rm A} \mid 
\hat{Q}\hat{M}^{(m)} \cdots \hat{Q}\hat{M}^{(1)}\mid \psi_{\rm A} \rangle 
/\langle \psi_{\rm A} \mid \hat{Q}\hat{M}^{(m-1)} \cdots 
\hat{Q}\hat{M}^{(1)}\mid \psi_{\rm A} \rangle $   
from 100 samples by the conventional SSS method with $\epsilon=0.01$, 
which we show by open circles. 
Another approximate state constructed by $887875$ basis states
($\mid \psi_{{\rm A}2} \rangle$) is also used in the constrained SSS method,
whose results from 100 samples with $\epsilon=0.01$ are plotted by filled squares.    
The dotted line in the figure shows  
the exact value of $Q$ obtained from the exact ground state energy\cite{Bernu}. 
Errors in the figure are statistical errors only.
Statistical errors for both filled marks are within marks. 
}
\label{fig:36site_a}
\end{center}
\end{figure}

\begin{figure}[h]
\begin{center}
\scalebox{0.45}{\includegraphics{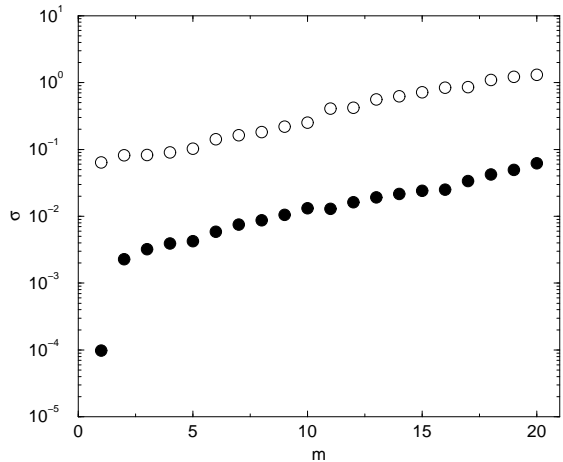}}
\caption{Standard deviations for ratios shown in Fig.~\ref{fig:36site_a},
which we obtain using the approximate state constructed by $333001$ basis states 
($\mid \psi_{{\rm A}1} \rangle$), from 100 samples with $\epsilon = 0.01$. 
Filled circles present $\sigma [R_A^{(m)}]$ obtained by the constrained SSS method.
Open circles are standard deviations for ratios obtained by the conventional SSS method.
}
\label{fig:36site_b}
\end{center}
\end{figure}

\begin{figure}[h]
\begin{center}
\scalebox{0.45}{\includegraphics{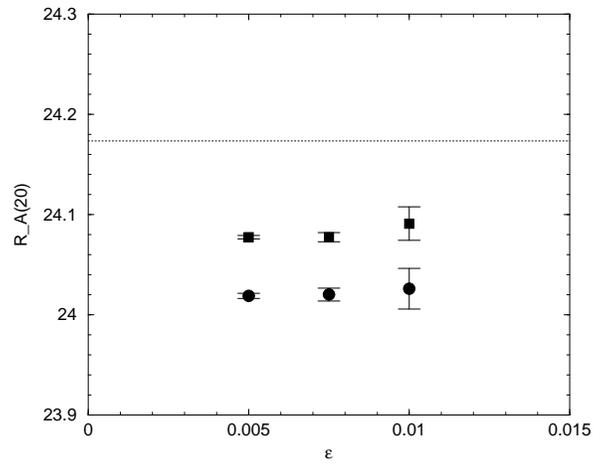}}
\caption{Ratios $R_{\rm A}^{(m_{\rm max})}$ with $m_{\rm max}=20$ 
versus $\epsilon$.
Circles (squares) present the results obtained from 10 samples with 
$\mid \psi_{{\rm A}1} \rangle$ ($\mid \psi_{{\rm A}2} \rangle$).
Errors shown in the figure are statistical errors only.
The standard deviations calculated from these $10$ samples are consistent with those 
shown in Fig.~\ref{fig:36site_a}, where the number of samples is $100$.
The dotted line indicates  
the exact value of $Q$ obtained from the exact ground state energy\cite{Bernu}. 
}
\label{fig:36site_d}
\end{center}
\end{figure}

\begin{figure}[h]
\begin{center}
\scalebox{0.45}{\includegraphics{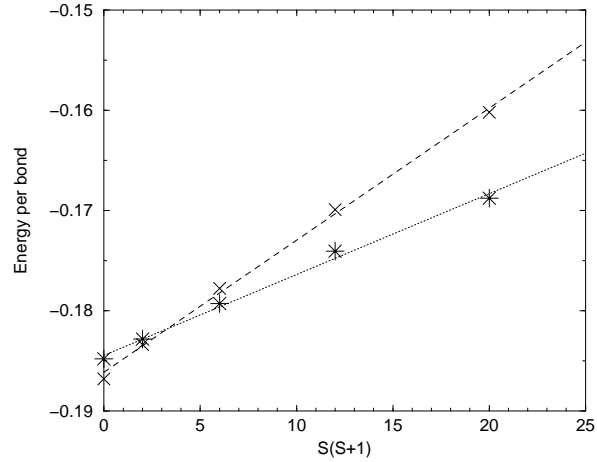}}
\caption{Energy per bond obtained by $\overline{E}_\kappa/(3N_s)$ $(\kappa=0,1,2,3,4)$   
on the 48-site lattice. The data are shown by asterisks, whose errors are within the marks. 
We also plot the 36-site data from ref.~\cite{Bernu} by crosses.  
The dotted (dashed) line is obtained by the least square fit assuming (\ref{ssb}) 
on the 48-site (36-site) lattice. 
}
\label{fig:48site_b}
\end{center}
\end{figure}

What can we say about the accuracy of our results?
Let us here estimate the systematic error which exists owing to the power method
with finite powers up to $m_{\rm max}$.  
On the 36-site lattice 
the exact ground state energy is known to be $E_{\rm g}= -20.1734$. 
The upper bound of the systematic error therefore can be estimated 
by $(E_{\rm ub}-E_{\rm g})/|E_{\rm g}|$ with an $E_{\rm ub}$,
an upper bound of $E$ given by $l - Q_{\rm lb}$ where 
$Q_{\rm lb}$ denotes the lower bound of $Q$. 
We see that $R_{\rm A}^{(m_{\rm max})}$ gives our best lower bound of $Q$,
because $R_{\rm A}^{(m)}$ increases as $m$ grows and 
$R_{\rm A}^{(m)} \rightarrow Q$ ($m \rightarrow \infty$) should hold.
Using $l-R_{\rm A}^{(20)}$ in Fig.~\ref{fig:36site_a} as 
$E_{\rm ub}$ we conclude that the systematic error is less than $0.8\%$ ($0.5\%$) 
for $\mid \psi_{{\rm A}1} \rangle$ ($\mid \psi_{{\rm A}2} \rangle$). 
On the 48-site lattice where no exact energy eigenvalue is known, we try to  
find a lower bound $E_{\rm lb}$ from our data 
using the upper bound $E_{\rm ub}$ of $E$. 
Note that this task is equivalent to find $Q_{\rm ub}$,
an upper bound of $Q$, because $E_{\rm lb}= l- Q_{\rm ub}$. 
In order to find an upper bound of $Q$ we carry out an empirical fit.
Details of this fit are described in Appendix C.
Then using $E_{\rm ub}$ and $E_{\rm lb}$ 
we evaluate the systematic error by $\left( E_{\rm ub}-E_{\rm lb}
\right) /|E_{\rm ub}|$
in each sector with $S_z=\kappa$.
Results from the data presented in Fig.~\ref{fig:48site_b} 
with $Q_{\rm lb}=R_{\rm A}^{(m_{\rm max})}
=R_{\rm A}^{(15)}$ are $0.6\%$, $0.9\%$, $0.8\%$, $3.2\%$ and $1.3\%$ for 
$S_z=0$, 1, 2, 3 and 4, respectively\footnote{ 
Similar analysis on the 36-site lattice yields the result that, with $m_{\rm max}=20$,
$\left( E_{\rm ub}-E_{\rm lb} \right) /|E_{\rm ub}|$ is  
$0.9\%$ ($0.4\%$) with $\mid \psi_{{\rm A}1} \rangle $ 
($\mid \psi_{{\rm A}2} \rangle$).}  

It would be the simplest way to impose only one constraint in each random choice 
operator ($K=1$) as we did in sections \ref{secexact} and \ref{secappx}.
Nevertheless one should remember that imposing more constraints in one random choice 
operator ($K > 1$) is also, at least theoretically, possible in the constrained SSS 
method as was discussed in section \ref{sec2}.
Although the $K > 1$ calculations might be numerically more difficult,    
further study for these cases is desired from practical point of view. 

Finally let us emphasize that the constrained SSS method, 
as well as the conventional SSS method, has no physical
bias since this method does not depend on any physical assumption
\footnote{Although in applications presented in this paper we assumed some symmetries 
to construct approximate states, these assumptions are not essential in the constrained 
SSS method itself. }. 
Therefore the method is applicable to numerical study of various systems.
Results obtained in this work on triangular lattices encourage us to numerically 
study spin systems on other lattices by means of this method.   

\vskip 1cm \noindent
{\Large {\bf Acknowledgement}}
\vskip 0.5cm

This work is supported by Grant-in-Aid for Scientific Research(C)
(19540398) from the Japan Society for the Promotion of Science.

\vskip 1cm \noindent
{\Large {\bf Appendix \ A}}

Here we describe how we determine the value of $l$ in the operator $\hat Q$ defined by 
(\ref{qdef}). Let us denote all eigenvalues of $\hat H$ by $E_{\rm {min}} ( < 0)$, $E_{\rm a}$, $E_{\rm b}$, $\cdots$, $E_{\rm {max}} ( > 0)$, where 
\begin{eqnarray*}
E_{\rm {min}} \leq E_{\rm a} \leq E_{\rm b} \leq \cdots \leq E_{\rm {max}} .
\end{eqnarray*}
It is easy to see $E_{\rm {max}}=3N_s/4$ for the $N_s$-site lattice. 
Since $\hat Q = l \hat I - \hat H$, eigenvalues of $\hat Q$ are then
\begin{eqnarray*}
l -E_{\rm {min}} \geq l- E_{\rm a} \geq l- E_{\rm b} \geq \cdots \geq l - E_{\rm {max}} .
\end{eqnarray*}

Let us consider a state $\mid \Psi \rangle $ and expand it using an orthonormal basis 
$\{ \mid \Psi_{\rm x} \rangle \}$ which is obtained from eigenfunctions of $\hat H$,
\begin{eqnarray*}
\mid \Psi \rangle = \sum _{\rm x} \mid \Psi_{\rm x} \rangle f_{\rm x} .
\end{eqnarray*}
Then we obtain 
\begin{eqnarray*}
\langle \Psi \mid \hat Q ^m \mid \Psi \rangle &=& \sum _{\rm x} f_{\rm x}^2 
\left( l-E_{\rm x} \right) ^m \\
&=& f_{\rm y}^2 \left( l-E_{\rm y} \right) ^m \{ 1+ \sum_{{\rm x} \neq {\rm y}} 
\frac{f_{\rm x}^2}{f_{\rm y}^2} 
\left( \frac{l-E_{\rm x}}{l-E_{\rm y}} \right) ^m \}
\\
&=& f_{\rm y}^2 \left( l-E_{\rm y} \right) ^m \{ 1+ \sum_{{\rm x} \neq {\rm y}} 
\frac{f_{\rm x}^2}{f_{\rm y}^2} \left(1-
\frac{E_{\rm x}-E_{\rm y}}{l-E_{\rm y}} \right) ^m \},
\end{eqnarray*}
where by suffix y we denote the term whose 
$| l-E_{\rm x} |$ is the largest among x with non-zero $f_{\rm x}$.
It is clear that the term $f_{\rm y}^2 \left( l-E_{\rm y} \right) ^m$ dominates as 
$m$ increases. 

When $\mid \Psi \rangle = \mid \psi_{\rm A} \rangle$ defined by (\ref{psia}) with 
$f_{\rm min}\mid \Psi_{\rm min} \rangle= w \mid \psi_{\rm E} \rangle$,  
the value of $l$ therefore should satisfy the condition  
\begin{eqnarray*}
| l-E_{\rm {min}} | > | l-E_{\rm {max}} | 
\end{eqnarray*}
in order for us to pick up the term with $f_{\rm min}^2 \left( l-E_{\rm min} \right) ^m$ in
$\langle \psi_{\rm A} \mid \hat Q ^m \mid \psi_{\rm A} \rangle$.
Limiting ourselves to the range $l < E_{\rm {max}}$ we thus see 
$l + | E_{\rm {min}} | > E_{\rm {max}} -l $, namely 
\begin{eqnarray*}
l > \frac{1}{2} \left( E_{\rm {max}}-| E_{\rm {min}} | \right) 
\end{eqnarray*}
should hold.
In most cases we have to use an upper bound of $ E_{\rm {min}}$,
$ E_{\rm {upper}}$, instead of $ E_{\rm {min}}$.
Then we determine the value of $l$ in the range
\begin{eqnarray*}
l > \frac{1}{2} \left( E_{\rm {max}}-| E_{\rm {upper}} | \right), 
\end{eqnarray*}
which includes the range $ l > \frac{1}{2} \left( E_{\rm {max}}-| E_{\rm {min}}| \right)$ 
for any $ E_{\rm {upper}} < 0$ because         
$E_{\rm {max}}-| E_{\rm {upper}} | > E_{\rm {max}}-| E_{\rm {min}} |$ holds. 

In choosing a value of $l$ which satisfy the above condition, 
one should also note that contributions from excited states
increase as $l$ increases because the dumping factors $\left(1-
\frac{E_{\rm x}-E_{\rm min}}{l-E_{\rm min}} \right) ^m $ decrease.

For the 16-site lattice, we use the exact value  
$E_{\rm {min}}= -8.5555$ to decide $l=2$. 
For 36-site and 48-site lattices we use $ E_{\rm {upper}}$ calculated from
$\langle \psi_{\rm A} \mid \hat H \mid \psi_{\rm A} \rangle$.
Values of $l$ we chose are summarized in the Table~2.
\vskip 1cm
\begin{table*}[h]
\begin{center}
\begin{tabular}{|c|c|c|c|c|} \hline
$N_s$ & $S_z$ & $E_{\rm {upper}}$ & $E_{\rm max} - |E_{\rm {upper}}|$ & $l$ \\ \hline
36 & 0 & $-19.57$ &  \ 7.43 & 4.0 \\ \cline{3-5}
   &   & $-19.81$ &  \ 7.19 & 4.0 \\ \hline
48 & 0 & $-26.41$ &  \ 9.59 &5.0 \\ \cline{2-5}
   & 1 & $-25.98$ & 10.02 & 5.2 \\ \cline{2-5}
   & 2 & $-25.48$ & 10.52 & 5.5 \\ \cline{2-5}
   & 3 & $-24.75$ & 11.25 & 5.8 \\ \cline{2-5}
   & 4 & $-24.09$ & 11.91 & 6.2 \\ \hline
\end{tabular}
\caption{Values of $l$ we use in the $S_z=\kappa$ sectors 
($\kappa=0$ for the 36-site lattice and $0 \leq \kappa \leq 4$ for the 48-site lattice), 
where $S_z$ is the $z$ component of the total spin $S$. 
$E_{\rm {upper}}$ is an upper bound for the lowest energy eigenvalue of the system, 
which we obtain from the expectation value of the Hamiltonian calculated with  
an approximate state, while
$E_{\rm max}$ denotes the maximum energy eigenvalue of the system. 
} 
\end{center}
\end{table*} 
 

\vskip 1cm \noindent
{\Large {\bf Appendix \ B}}

Here we explain how we obtain approximate states $\mid \psi_{\rm A} \rangle$'s 
on the $36$-site and $48$-site lattices.  
The method is essentially the same as the one given in ref.~\cite{MMtri} 
for the 36-site triangular lattice.
The only difference is that we include here as many degenerate Ising-like configurations
as possible in an initial trial state. This improvement
comes from Wannier's rigorous proof\cite{Wannier} which claims 
that a classical antiferromagnetic Ising system, 
namely the spin system at zero temperature $T=0$, on an $N_s$-site triangular lattice 
is heavily degenerated for its minimum energy $ -N_s/4$.  

For our numerical work in this paper, 
we employ one basis on the 36-site lattice and five bases on the 48-site lattice corresponding to values of $S_z$.
First we comment on these bases together with brief descriptions 
for the transformation symmetries we impose on them. 
Then we show procedures to create an approximate state for numerical studies.

Each state $\mid j \rangle $ in our basis states $\{ \mid j \rangle \}$ is
represented by linear combinations of states 
$\mid s_1, s_2, \cdots, s_{N_s} \rangle$ with $s_n = +1/2$ or $-1/2$ 
($n=1,2,\cdots,N_s$), which indicates that the $z$-component of the spin 
on the $n$-th site is $+1/2$ or $-1/2$, respectively.
They belong to one of $2N_s+1$ sectors according to the value of $S_z=\sum_{n=1}^{N_s} s_n$, 
which is the $z$-component of the total spin $S$ of the system.

The basis states to form the approximate state for the ground state 
on the $36$-site lattice should have symmetries found in ref. \cite{Bernu}.
Following them we construct our basis states with $S_z=0$ so that they have 
translational symmetries for the zero momentum $(0,0)$ 
and even under the $2\pi/3$ rotation and the reflection.
Each space group which consists of the translation group and the point group 
has then $432(=12 \times 36)$ elements.

For the model on the 48-site lattice we assume each basis state with a 
given value of $S_z$ has   
the same symmetries as those of the exact state for the $N_s=36$ lattice\cite{Bernu}
whose energy eigenvalue is the lowest in the corresponding sector.
Namely, all basis states in both bases 
have translational symmetries for the zero momentum $(0,0)$ 
and even under the $\pi$ rotation, the $2\pi/3$ rotation and the reflection. 
The basis states in the basis used for $S_z=0, 2$ and $4$ calculations also have the even $\pi$ 
rotational symmetry, while those in another basis prepared for $S_z=1$ and $3$ 
have the odd one.
Both space groups have $12 N_s = 576$ elements. 

Now we start to calculate $ \mid \psi_{\rm A} \rangle $ in each sector with a fixed 
value of $S_z$, using an appropriate basis among those stated above. 

The first stage is to find as many degenerate states for the eigenvalue $-N_s/4$ as possible.
Using the conventional Monte Carlo method at low temperature($T=0.5$), where
the classical energy is used as the Boltzmann weight, we pick up $N_{\rm it} \sim 10^4$ 
states to fulfil the condition 
$ \langle j \mid \hat{H} \mid j \rangle = -N_s/4$ and $S_z=\kappa$ $(0 \leq \kappa \leq 4)$,
where the Hamiltonian $\hat H $ is given by (\ref{hamil}). 
With this $N_{\rm it}$ basis states 
we calculate the eigenstate $ \mid \Psi_{\rm t} \rangle $ for the lowest energy eigenvalue 
within this partial Hilbert space by means of the conventional exact diagonalization. 

The next stage to calculate an approximate state $\mid \psi_{\rm A} \rangle$ 
is to repeat following procedures, starting from the initial $ \mid \Psi_{\rm t} \rangle $
obtained above, 
until the expectation value for $\hat H$ does not change beyond our criteria.

\begin{enumerate}
\item
Extend the partial Hilbert space given by $ \mid \Psi_{\rm t} \rangle  $ 
through operations of a few $\hat{H}$'s to $\mid \Psi_{\rm t} \rangle$, 
until the available computer memory is exhausted. 
The maximum number of basis states we can permit is about $ 10^8$.
\item
Within the Hilbert space determined in (i), pursue the state 
$\mid  \Psi_{\rm t'} \rangle$ with which 
$\langle \Psi_{\rm t'} \mid  \hat{H} \mid  \Psi_{\rm t'} \rangle$ is as low as possible.

\item
If the change of the obtained value  
$\langle \Psi_{\rm t'} \mid  \hat{H} \mid  \Psi_{\rm t'} \rangle$
is in the range of five decimal digits,   
employ $\mid \Psi_{\rm t'} \rangle$ as the approximate state $\mid \psi_{\rm A} \rangle $. 
Otherwise, form a state $\mid \Psi_{\rm t''} \rangle $ 
by keeping a few percent of the basis states 
whose coefficients in the expansion of $\mid  \Psi_{\rm t'} \rangle $ are relatively large
and replace $\mid  \Psi_{\rm t} \rangle $ by $ \mid \Psi_{\rm t''} \rangle$ to 
proceed with (i), (ii) and (iii) once more.
\end{enumerate}

\vskip 1cm \noindent
{\Large {\bf Appendix \ C}}

\begin{figure}[h]
\begin{center}
\scalebox{0.45}{\includegraphics{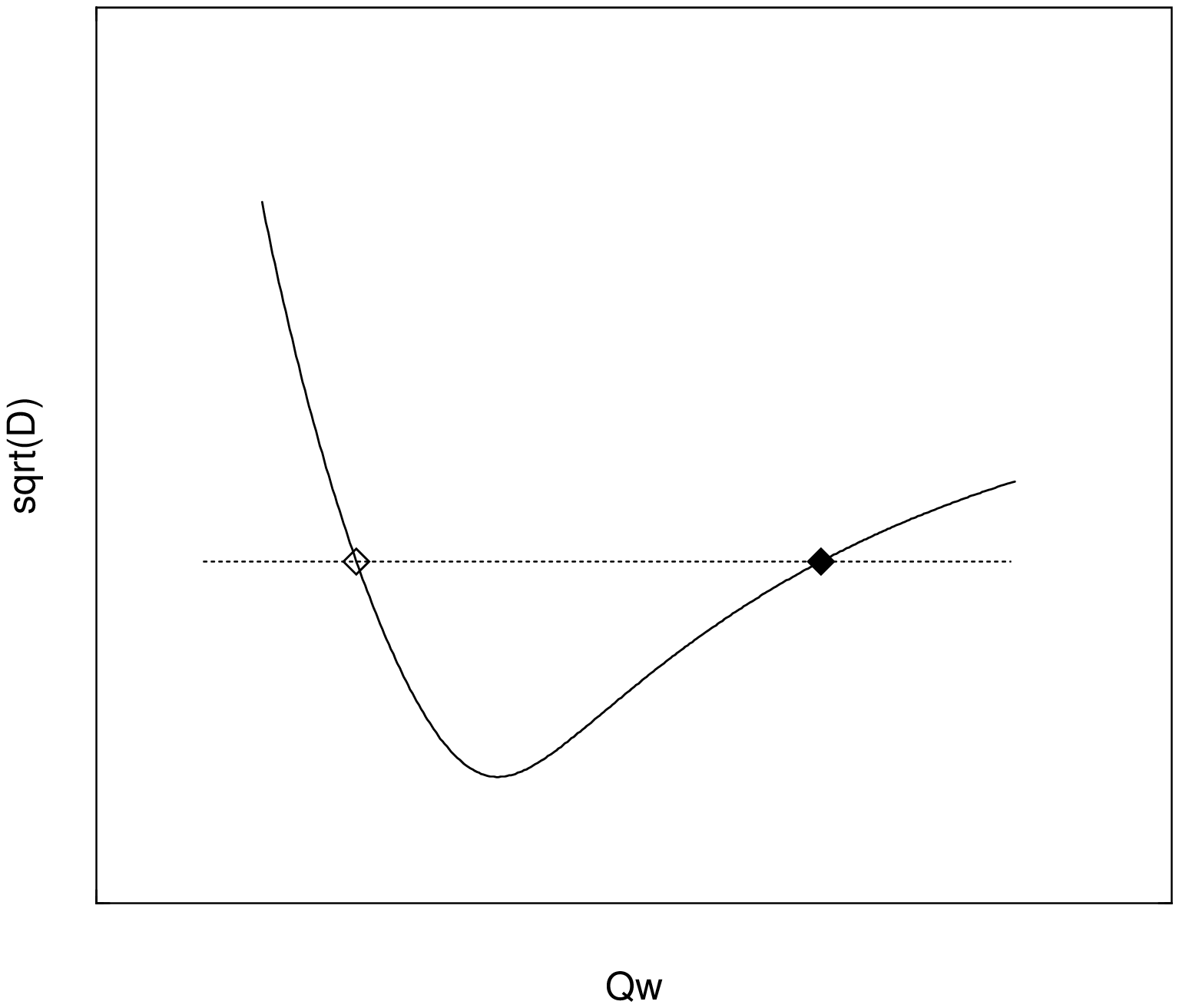}}
\caption{Typical plots of 
$\sqrt{D(m_{\rm max}, Q_{\rm w},\underline{q}_0, \underline{\alpha})}$,
where $D(m_{\rm max}, Q_{\rm w},q_0, \alpha) $ is
defined by (\ref{defdd}), as a function of $Q_{\rm w}$.  
Using this figure we can find $Q_{\rm ub}$, which is indicated by a filled diamond, 
from a given value of $Q_{\rm lb}$ indicated by an open diamond.
The dotted horizontal line is to guide eyes.
}
\label{fig:sqrtd_vs_qw}
\end{center}
\end{figure}

Here we explain how we evaluate the lower bound of the eigenvalue of 
$\hat Q = l \hat I -\hat H$, which we denote by $Q_{\rm lb}$, 
from the data $R_{\rm A}^{(m)}$ $(m=1,2,\cdots,m_{\rm max})$ and 
a given value of the upper bound of the eigenvalue of $\hat Q$ denoted by $Q_{\rm ub}$. 
We employ an empirical formula we presented in a previous paper\cite{MM1}, which is
\begin{equation}
F(m,Q_{\rm w},q_0,\alpha) \equiv 
Q_{\rm w}^m \left( q_0 + \frac {q_1}{m+\alpha+1} \right), \ \ \ q_1=(\alpha+1)(1-q_0).
\label{deff}
\end{equation}
The quantity we measure is $\sqrt{D(m_{\rm max},Q_{\rm w},q_0,\alpha)}$, 
\begin{equation}
D(m_{\rm max},Q_{\rm w},q_0,\alpha) \equiv \sum _{m=1}^{m_{\rm max}} \left[ 
1- \frac{\langle \psi_{\rm A} \mid \phi_{\rm A}^{(m)} 
\rangle}{F(m,Q_{\rm w},q_0,\alpha)} \right] ^2 ,
\label{defdd}
\end{equation}
where values of $ \langle \psi_{\rm A} \mid \phi_{\rm A}^{(m)}\rangle$ 
$(m=1,2,\cdots,m_{\rm max})$ are calculated from the data $R_{\rm A}^{(m)}$,
\begin{equation}
 \langle \psi_{\rm A} \mid \phi_{\rm A}^{(m)} \rangle 
= \prod _{n=0}^{m-1} R_{\rm A}^{(n)}, \ \ \ R_{\rm A}^{(0)} \equiv 1 .
\label{pampa}
\end{equation} 
Changing $Q_{\rm w}$ around $Q_{\rm lb}$ which is given by the data 
$R_{\rm A}^{(m_{\rm max})}$, we look for values of parameters $q_0$ and $\alpha$,
say $\underline{q}_0$ and $\underline{\alpha}$, 
so that $D(m_{\rm max},Q_{\rm w},\underline{q}_0,\underline{\alpha}) $ gives a local minimum 
for the given value of $Q_{\rm w}$.  
Fig.~\ref{fig:sqrtd_vs_qw} shows a typical result for 
$\sqrt { D(m_{\rm max},Q_{\rm w},\underline{q}_0,\underline{\alpha}) } $ 
as a function of $Q_{\rm w}$. 
By requesting that the value of $\sqrt{D}$ at $Q_{\rm w}=Q_{\rm ub}$ should be equal to 
the value at $Q_{\rm w}=Q_{\rm lb}$, we obtain a upper bound $Q_{\rm ub}$ shown in the figure.


\vskip 0.5cm

\begin{thebibliography}{23}

\bibitem{book1}
Hatano N and Suzuki M 1993
{\it Quantum Monte Carlo Methods in Condensed Matter Physics}
ed M Suzuki (Singapore: World Scientific) p~13 

\bibitem{book2}
De Raedt H and von der Linden W 1995  
{\it The Monte Carlo Method in Condensed Matter Physics}
ed K Binder (Berlin: Springer) p~249

\bibitem{rev}
Richter J, Schulenburg J and Honecker A 2004 {\it Quantum Magnetism 
(Lecture notes in physics {\bf 645})}
ed U Schollw\"{o}ck, J Richter, D J J Farnell and R F Bishop 
(Berlin/Heidelberg: Springer-Verlag)


\bibitem{ccm1}
Farnell D J J, Bishop R F and Gernoth K A 2001 Phys. Rev. B {\bf 63} 220402

\bibitem{ccm2}
Kr\"{u}ger S E, Darradi R, Richter J, Farnell D J J 2006
Phys. Rev. B {\bf 73} 094404 


\bibitem{Sorella}
Sorella S 2001 Phys. Rev. B {\bf 64} 024512

\bibitem{fixednode2}
van Bemmel H J M, ten Haaf D F B, van Saarloos W, van Leeuwen J M J
and An G 1994 Phys. Rev. Lett. {\bf 72} 2442

\bibitem{fixednode}
ten Haaf D F B, van Bemmel H J M, van Leeuwen J M J,
van Saarloos W and Ceperley D M 1995 Phys. Rev. B {\bf 51} 13039

\bibitem{Hene}
Henelius P 1999 Phys. Rev. B {\bf 60} 9561

\bibitem{Mae}
Maeshima N, Hieida Y, Akutsu Y, Nishino T and Okunishi K 2001 Phys. Rev. E 
{\bf 64} 016705

\bibitem{White3}
White S R and Chernyshev A L 2007 Phys. Rev. Lett. {\bf 99} 127004

\bibitem{White1}
White S R 1992 Phys. Rev. Lett. {\bf 69} 2863

\bibitem{White2}
White S R 1993 Phys. Rev. B {\bf 48} 10345

\bibitem{IK}
Imada M and Kashima T 2000 J. Phys. Soc. Japan {\bf 69} 2723


\bibitem{MM1}
Munehisa T and Munehisa Y 2003 J. Phys. Soc. Japan {\bf 72} 2759

\bibitem{MMss} 
Munehisa T and Munehisa Y 2004 J. Phys. Soc. Japan {\bf 73} 340

\bibitem{MM2} 
Munehisa T and Munehisa Y 2004 J. Phys. Soc. Japan {\bf 73} 2245

\bibitem{MM3} 
Munehisa T and Munehisa Y 2004 Numerical study for an equilibrium in the
recursive stochastic state selection method {\it Preprint} cond-mat/0403626 

\bibitem{MMtri}
Munehisa T and Munehisa Y 2006 J. Phys. : Condens. Matter {\bf 18} 2327

\bibitem{MMeq} 
Munehisa T and Munehisa Y 2007 J. Phys. : Condens. Matter {\bf 19} 196202


\bibitem{Singh}
Singh R R P and Huse D A 1992 Phys. Rev. Lett. {\bf 68} 1766

\bibitem{Leung}
Leung P W and Runge K J 1993 Phys. Rev. B {\bf 47} 5861

\bibitem{Sindz}
Sindzingre P, Lecheminant P and Lhuillier C 1994 Phys. Rev. B {\bf 50} 3108

\bibitem{Bernu}
Bernu B, Lecheminant P, Lhuillier C and Pierre L 1994 Phys. Rev.  B {\bf 50} 10048

\bibitem{Boninsegni}
Boninsegni M 1995 Phys. Rev. B {\bf 52} 15304

\bibitem{Capri}
Capriotti L, Trumper A E and Sorella S 1999 Phys. Rev. Lett. {\bf 82} 3899

\bibitem{Weihong}
Zheng W, McKenzie R H and Singh R R P 1999 Phys. Rev. B {\bf 59} 14367

\bibitem{Trumper}
Trumper A E, Capriotti L and Sorella S 2000 Phys. Rev. B {\bf 61} 11529

\bibitem{Sorella2}
Arrachea L, Capriotti L and Sorella S 2004 Phys. Rev. B {\bf 69} 224414

\bibitem{Sorella1}
Yunoki S and Sorella S 2006 Phys. Rev. B {\bf 74} 014408


\bibitem{Fendley}
Fendley P, Moessner R and Sondhi S L 2002 Phys. Rev. B {\bf 66} 214513

\bibitem{Oleg}
Starykh O A, Chubukov A V and Abanov A G 2006 Phys. Rev. B {\bf 74} 180403

\bibitem{Zheng}
Zheng W, Fjaerestad J O, Singh R R P, McKenzie R H and Coldea R 2006 Phys. Rev. B {\bf 74} 224420

\bibitem{Neuberger}
Neuberger H, Ziman T 1989 Phys. Rev. B {\bf 39} 2608

\bibitem{Bernu2}
Bernu B, Lhuillier C and Pierre L 1992 Phys. Rev. Lett. {\bf 69} 2590

\bibitem{Wannier}
Wannier G H 1950 Phys. Rev. {\bf 79} 357

\end{thebibliography}
\end{document}